\documentclass[pra,final,twocolumn,showpacs,aps]{revtex4-2}
\usepackage{multirow}
\usepackage{graphicx}
\usepackage{amssymb}
\usepackage{amsmath}
\usepackage{color}
\usepackage{xcolor}
\usepackage{multirow}
\usepackage{afterpage}
\usepackage{hyperref}
\usepackage{float}
\usepackage{ulem}

\begin{document}
\title[Qt in DBEC]{Study of quantum turbulence by vortex-antivortex dynamics in dipolar BECs
}
\author{S. Sabari\email{ssabari01@gmail.com}, 
and Lauro Tomio\email{lauro.tomio@unesp.br}}
\affiliation{{Instituto de F\'\i sica Te\'{o}rica, Universidade 
Estadual Paulista, 01140-070 S\~{a}o Paulo, SP, Brazil}
}\date{\today}

\begin{abstract}
We investigate vortex nucleation and dynamics in a dipolar Bose-Einstein condensate
stirred by a rotating Gaussian obstacle. Two stirring protocols are compared: continuous
stirring with constant amplitude (Case 1), and obstacle removal at $t = 15$~{ms} with
linear amplitude ramp-down (Case 2). In both cases, a smooth angular velocity ramp
($t_{\text{acc}} = 3$~{ms}) suppresses spurious phonon excitations. {The first vortex pairs exit the obstacle at $t \sim 3$~ms, and by
$t \sim 8$--$10$~ms the individual vortices and antivortices are fully resolved
in the phase maps.} Case 1 yields a {vortex population that grows
up to $t \sim 40$~ms and then roughly saturates}, 
while Case 2 yields {a population that peaks around $t \sim 15$--$25$~ms and then
decreases.}
Remarkably, both protocols lead to {local triangular-like ordering
in selected regions of the condensate} by $t \sim 50$~{ms}, exhibiting clear dipolar
signatures: elliptical distortion and elongated vortex cores along the polarization axis.
The mixed vortex-antivortex population remains stable up to $100$~{ms} with no
observable annihilation, indicating that dipolar interactions strongly suppress vortex
decay. Kinetic energy decomposition confirms that incompressible (vortex) energy dominates
at late times. These results establish a controlled platform for vortex
studies in dipolar superfluids and provide benchmarks for future experiments on quantum
turbulence with long-range anisotropic interactions.
\end{abstract}

\maketitle

\section{Introduction}
\label{sec1}

Dipole-dipole interactions (DDI) in ultracold gases have opened new avenues for exploring
quantum many-body phenomena beyond short-range contact interactions. Unlike isotropic 
$s$-wave scattering, DDI are anisotropic and long-range, leading to novel phases such as
roton instabilities, supersolids, and quantum droplets \cite{Baranov2008,Lahaye2009}.
Experimental realizations of dipolar Bose-Einstein condensates (BECs) with chromium
\cite{Griesmaier2006}, dysprosium \cite{Lu2011}, erbium \cite{Aikawa2012}, and more
recently europium \cite{2022Miyazawa}, have transformed these theoretical ideas into
testable realities.
A defining feature of dipolar BECs is the tunability of both contact and dipolar
interactions. The $s$-wave scattering length can be adjusted via Feshbach resonances 
\cite{1998Inouye}, while the magnitude and sign of the DDI can be controlled by rotating
the external magnetic field relative to the atomic dipole moments 
\cite{Giovanazzi2002,Sabari2015,Sabari2018a,Sabari2022}. This dual tunability makes 
dipolar condensates ideal platforms for studying fundamental aspects of superfluidity,
including quantized vortices and quantum turbulence.

Quantized vortices are topological defects that play a central role in superfluid 
physics. They have been studied extensively in nondipolar BECs using various nucleation
techniques: rotating the trapping potential \cite{2001Madison,Tamil2019}, stirring with a
focused laser beam \cite{Raman1999,Raman2001}, moving a condensate past a static defect
\cite{Henn2009}, and rapid quench protocols \cite{Lamporesi2013,Kishor2025,Lauro2024,Lauro2026}. 
In parallel, the study of quantum turbulence-the complex, disordered motion of quantized
vortex lines-has developed into a rich field, bridging concepts from classical fluid
dynamics with the discrete, quantum nature of vorticity in superfluids. After many years
of research with superfluid helium systems, quantum turbulence became a well-established 
field for investigation \cite{Barenghi2001}, with particular interest in cold-atom 
physics given the available experimental possibilities \cite{Pethick2008}.

Despite this progress, most studies of vortex dynamics and quantum turbulence have
focused on nondipolar atomic species. However, significant contributions have emerged
from investigations of dipolar BECs, particularly regarding vortex formation, hidden
vortices, and vortex-lattice structures under rotation 
\cite{Sabari2017,Sabari2026,Sabari2025}. The influence of anisotropic, long-range DDI on
vortex nucleation and dynamics has been explored in various contexts, including the 
effects of oscillating Gaussian obstacles \cite{Sabari2018}, vortex-antivortex pair
production in perturbed dipolar condensates \cite{Sabari2024cnf,Sabari2024}, and the role
of two- and three-body interactions in rotating dipolar BECs \cite{Sabari2018cnf}.
Hydrodynamic aspects of vortices and solitons in dipolar BECs have also been investigated \cite{Sabari2017cnf,Sabari2016cnf}.
Controlled vortex nucleation using a moving or rotating obstacle, a standard technique in
nondipolar BECs, has received focused attention in the dipolar context. Notable studies 
include the characterization of turbulence in a dipolar Bose gas condensing from a highly
non-equilibrium thermal state without external forcing \cite{2022Klaus}, and our previous
works \cite{Sabari2024cnf,Sabari2024}, where we investigated vortex nucleation and 
quantum turbulence in dipolar BECs using two distinct stirring protocols. Additionally,
the dynamics of vortices in dipolar BECs under oscillating potentials has been examined
\cite{Sabari2018}.

In our previous studies \cite{Sabari2017,Sabari2024cnf,Sabari2024,Sabari2017cnf}, we
demonstrated that vortex dipoles can be nucleated in a dipolar BEC by stirring 
mechanisms, and we provided evidence for the onset of quantum turbulence under certain
parameter regimes. Those works focused primarily on the role of the dipolar interaction
strength, tuned by varying the angle of the external magnetic field, as well as the 
effects of two- and three-body interactions and oscillating potentials. While 
informative, the stirring protocols employed in those studies did not incorporate smooth
ramping of the obstacle motion, potentially introducing spurious phonon excitations that
could obscure the intrinsic vortex dynamics.

The present work serves as a complementary extension to our previous study. Here, we 
refine the stirring protocol by introducing a smooth angular velocity ramp at the 
beginning of the rotation, which minimizes unwanted excitations and allows us to isolate
the genuine vortex dynamics induced by the moving obstacle. Furthermore, we compare two
distinct time-dependent scenarios: continuous stirring with constant obstacle amplitude,
and a protocol in which the obstacle is smoothly removed after a finite time, allowing 
the vortex ensemble to evolve freely. These two cases provide complementary insights: 
continuous stirring mimics a steady-state driven system, while obstacle removal probes
the intrinsic relaxation and self-organization of the vortex ensemble in the absence of
external forcing. The observed self-organization into a stable, dipolar-modified vortex
lattice provides a benchmark for future experimental studies using dysprosium, erbium, or
europium condensates. Moreover, the long-lived vortex arrays demonstrated here may serve
as initial states for investigating quantum turbulence in anisotropic superfluids, a 
regime that remains largely unexplored {\cite{2026Zhu,2026Mohapatra,2025Jha,2024Prasad}}.

{Recent studies have further advanced this direction: strong and weak wave
turbulence regimes have been characterized in BECs
\cite{2026Zhu}, vortex-pair dynamics have been explored in three-dimensional
homogeneous dipolar superfluids \cite{2024Prasad}, energy spectra and fluxes have
been computed for two-dimensional turbulent quantum droplets \cite{2025Jha}, and
optical-lattice techniques have been proposed to eliminate vortices from condensates
\cite{2026Mohapatra}. These developments motivate the controlled stirring protocols
investigated in the present work.}

The paper is structured as follows. In Sec.~\ref{sec2}, we present the theoretical
model, including the extended Gross-Pitaevskii equation, the dipolar interaction
potential, and the detailed time-dependent stirring protocols.
Section~\ref{sec3} presents the main results, showing the time evolution of the
density, phase, kinetic energies, and their decomposition into compressible and incompressible components. It also discusses
the incompressible kinetic energy spectrum and its adherence to the Kolmogorov
scaling law $k^{-5/3}$ for two types of stirring protocols. Finally, Sec.~\ref{sec4} summarizes our conclusions and
outlines future directions for studying quantum turbulence in dipolar superfluids.

\section{Gross-Pitaevskii stirring model with dipole-dipole interaction}
\label{sec2}
In the mean-field approximation of a dilute dipolar BEC of atoms with mass $m$, 
assuming two-body contact interactions between the $N$ atoms, the following 
three-dimensional (3D) time-dependent Gross-Pitaevskii (GP) equation is obtained 
for the wave function $\Psi({\mathbf r},t)$ normalized to one
\cite{Lahaye2009,Sabari2024}: 
{\small
\begin{align}
{\rm i}\hbar\frac{\partial \Psi({\mathbf r},t)}{\partial t}& =
\left(-\frac{\hbar^2}{2m}\nabla^2+ V({\mathbf r},t) + g_{3D}
 \left\vert \Psi({\mathbf r},t)\right\vert^2 \right)\Psi({\mathbf r},t)
 \notag \\ 
 &+N \int d^3{\mathbf r}' U_{\mathrm{dd}}({\mathbf  r}-{\mathbf r}')
 \left\vert\Psi({\mathbf r}',t)\right\vert^2 
 \Psi({\mathbf r},t), \label{eqn:dgpe}
\end{align}
}where $g_{3D}\equiv 4\pi\hbar^2 a_sN/m$ (with $a_s$ being the $s-$wave scattering length) is the cubic nonlinear contact parameter. 
The confining trap potential $V({\mathbf r},t)$ is defined as, 
\begin{eqnarray}
V({\mathbf r},t) = \frac{m\omega_\rho^2}{2}\left[(x^2+y^2)+
\lambda z^2\right]+V_G(x,y,t),
\label{trap}
\end{eqnarray}
where 
{\small\begin{eqnarray}\label{VG}
V_{G}& = &A(t) \exp\left(-\frac{\left[x-x_0(t)\right]^2+
\left[y-y_0(t)\right]^2}{2\sigma^2}\right) \label{eqn:Obstacle}
\end{eqnarray}
} {is the rotating Gaussian obstacle potential,} with $x_0(t)=r_0 \cos\theta(t)$, and $y_0(t)=r_0 \sin\theta(t)$. 

Following the procedure discussed in Ref.~\cite{Sabari2024}, we obtain the effective 
2D equation for the pancake-shaped dipolar BEC as
\begin{eqnarray}
 {\rm i}\frac{\partial \psi}{\partial t}&=&\left\{
 -\frac{1}{2}\nabla_{\rho}^2+\frac{\rho^2}{2}+ V_{G} +g_{s}|\psi|^2 \right.
\nonumber\\
&+&\left. g_{dd}\int \frac{d^2k_{\rho}}{4\pi^2}
e^{i \mathbf{k}_{\rho}.\tilde{\rho}}\tilde{n}(\mathbf k_{\rho})
\widetilde{{V}}^{(d)}({\bf k_\rho}) \right\} \psi,
\label{gpe_scaled}
\end{eqnarray}
where $\psi\equiv\psi(\boldsymbol{\rho}, t)$ is the 2D wave function, normalized 
to one. To solve Eq.~(\ref{gpe_scaled}) and investigate vortex nucleation and pair
production in the dipolar BEC, we employ the split-step Crank-Nicolson method combined
with the Fast Fourier Transform approach~\cite{Sabari2024}. The numerical grid consists
of $256 \times 256$ points with spacing $\Delta x = \Delta y = 0.1\,\ell_\rho$ and time
step $\Delta t = 0.0001\,\omega_\rho^{-1}$. Convergence was verified by doubling the grid resolution.
The following parameters are fixed throughout this study: $a_s = 100\,a_0$, 
$a_{dd} = 131\,a_0$, $N = 2.5 \times 10^4$, $A_0 = 300\,\hbar\omega_\rho$, 
$r_0 = 6.5\,\ell_\rho$, and $\sigma = 0.7\,\ell_\rho$, where $a_0 = 5.29177 \times 10^{-11}$ m
is the Bohr radius and $\ell_\rho = 1\;\mu\text{m}$ is the radial harmonic oscillator length.
The obstacle orbital radius $r_0$ is chosen to lie at an intermediate distance between
the center and the boundary of the condensate, as the critical velocity for {vortex dipole}
production scales as $v_c = r_0 \nu_c$. When $r_0 \to 0$, no {vortex dipoles} are produced.

\section{Results and Discussion}
\label{sec3}

We now present the time evolution of the density, phase, kinetic energies ($E_K$), 
and their decomposition into compressible ($E_K^{(c)}$) and incompressible ($E_K^{(i)}$)
components for both stirring protocols. The incompressible kinetic energy spectrum 
and its adherence to the Kolmogorov scaling law $k^{-5/3}$ are also analyzed.
The rotating obstacle potential, given by Eq.~(\ref{eqn:Obstacle}), has orbital radius
$r_0 = 6.5\,\ell_\rho$, width $\sigma = 0.7\,\ell_\rho$, and final angular velocity 
$\nu_0 = 1.0$ rad/ms. 
\begin{figure}[htbp]
\centering
\includegraphics[width=\columnwidth]{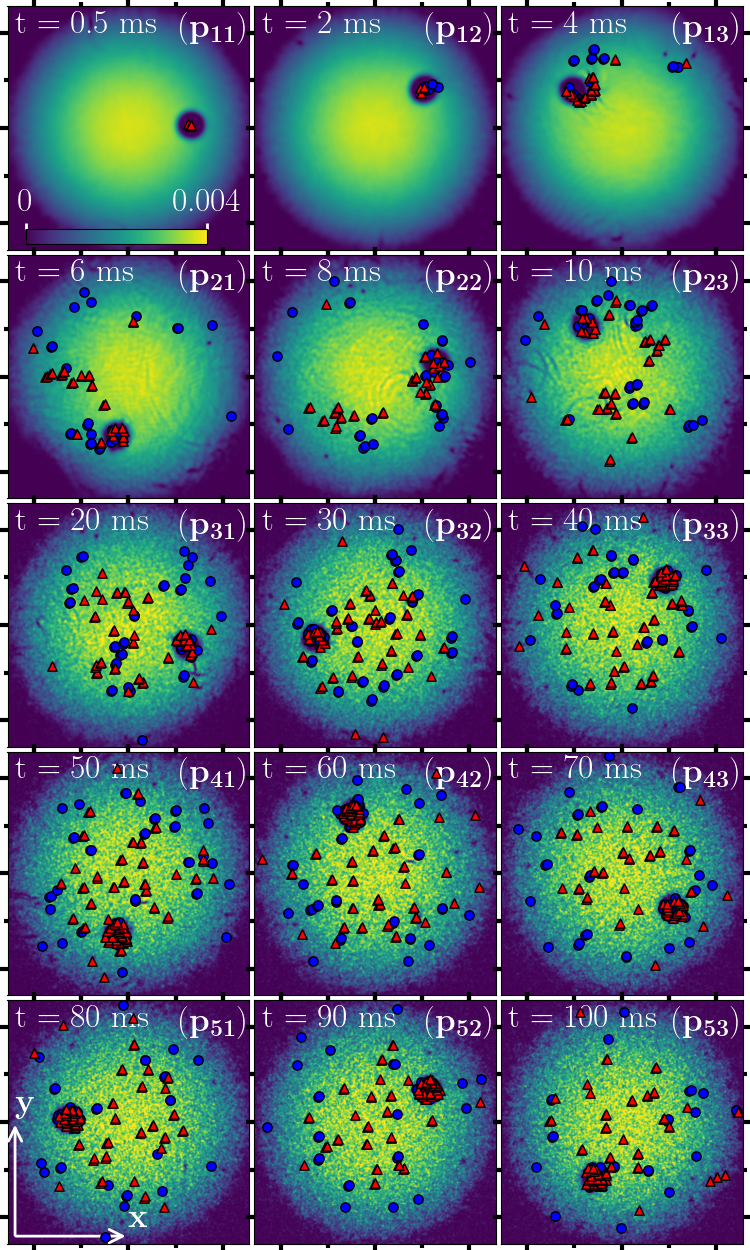}
\caption{
(Color online) Column-density evolution for a dipolar BEC continuously stirred by a
rotating Gaussian obstacle (Case 1). Parameters: $N=2.5\times10^4$, $a_s=100\,a_0$,
$a_{dd}=131\,a_0$, $A_0=300\,\hbar\omega_\rho$, $\sigma=0.7\,\ell_\rho$,
$r_0=6.5\,\ell_\rho$, $\nu_0=1.0$~rad/ms, and $t_{\text{acc}}=3$~ms. The
$5\times3$ panels correspond to
$t = 0.5, 2, 4, 6, 8, 10, 20, 30, 40, 50, 60, 70, 80, 90$, and $100$~ms.
Vortices appear as dark density holes. {The first vortex pairs exit the
obstacle at $t \sim 3$~ms, becoming visible in the $t=4$~ms panel.} {By
$t\sim50$~ms, local triangular-like ordering develops in selected regions of the
condensate, rather than a globally ordered lattice.} {Insets in the
$t=40$~ms and $t=50$~ms panels show the vortex positions together with
nearest-neighbor connectivity lines, highlighting the emergence of local
sixfold-coordinated regions.}}
\label{fig:dens_case1}
\end{figure}
\begin{figure}[htbp]
\centering
\includegraphics[width=\columnwidth]{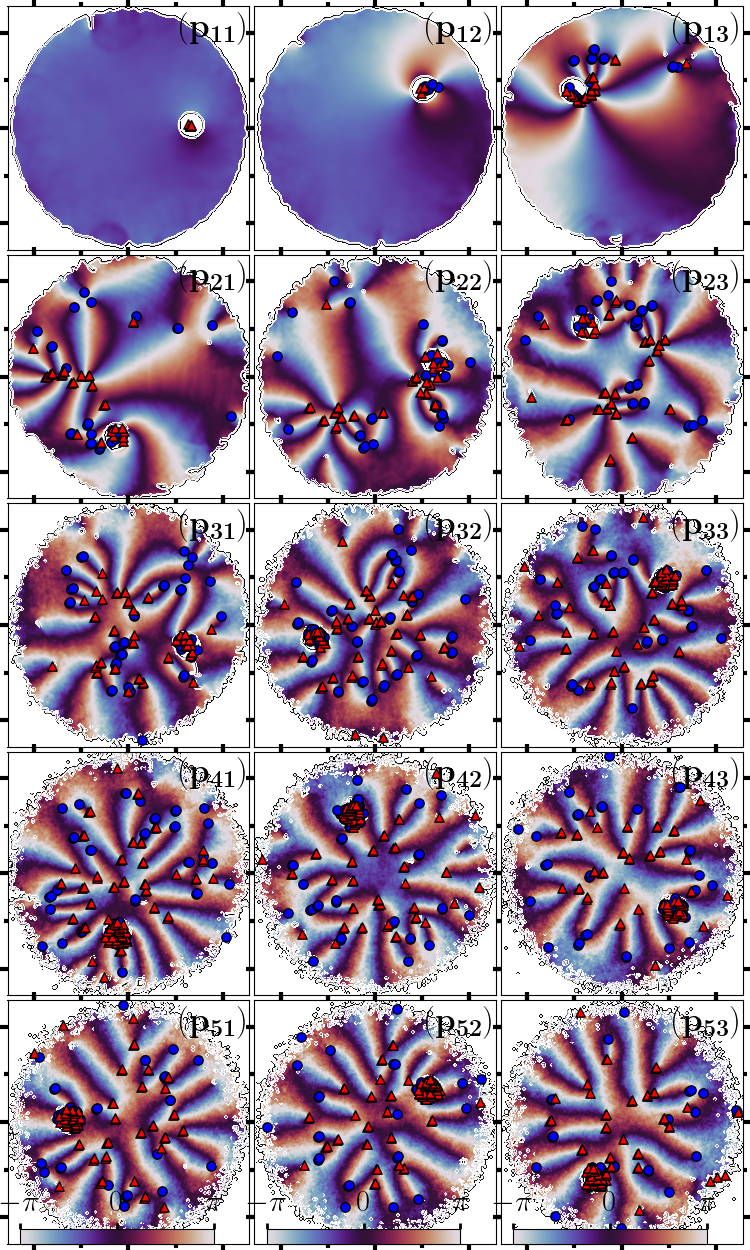}
\caption{
(Color online) Phase evolution corresponding to Fig.~\ref{fig:dens_case1} (Case 1).
Vortices appear as $2\pi$ phase dislocations. The phase maps confirm nucleation at
$t\sim8$-$10$ ms, a disordered phase at $20$-$40$ ms, and gradual ordering into a
triangular vortex lattice by $t\sim50$ ms.}
\label{fig:phase_case1}
\end{figure}
To avoid spurious phonon excitations,
the angular velocity is ramped linearly from $0$ to $\nu_0$ over an acceleration
time $t_{\text{acc}} = 3$~ms, after which it is held constant at $\nu_0$; the
corresponding angular displacement $\theta(t)$ is obtained by integrating the
angular velocity over time. Two cases are considered. In \textbf{Case 1
(continuous stirring)}, the obstacle amplitude remains constant at
$A(t) = A_0 = 300\,\hbar\omega_\rho$ for all $t$. In \textbf{Case 2 (obstacle
removal)}, the amplitude ramps linearly to zero over $t_{\text{remove}} = 15$~ms,
i.e., $A(t) = A_0(1 - t/t_{\text{remove}})$ for $0 \le t \le t_{\text{remove}}$,
and $A(t) = 0$ thereafter.

\subsection{Case 1: Continuous stirring}

Figures~\ref{fig:dens_case1} and \ref{fig:phase_case1} show the column density and
phase evolution for Case 1. The 15 panels ($5\times3$ grid) correspond to 
$t =$ 0.5, 2, 4, 6, 8, 10, 20, 30, 40, 50, 60, 70, 80, 90, 100~ms. At $t = 0.5$~ms
($p_{11}$), the vortex dipole begins to emerge but remains 
inside the rotating obstacle. {At $t = 4$~ms ($p_{13}$), the first vortex
pairs have already exited the obstacle and appear as distinct dark holes in the
density with $2\pi$ phase windings; this marks the onset of vortex-pair emission
at $t \sim 3$~ms.} At $t = 6$~ms ($p_{21}$), additional vortex dipoles are nucleated.
At $t = 8$~ms ($p_{22}$) and $t = 10$~ms ($p_{23}$), both vortices ($+2\pi$) and
antivortices ($-2\pi$) proliferate. Between $t = 20$ and $40$~ms ($p_{31}$--$p_{33}$),
the vortex and antivortex populations increase, but their spatial distribution remains
irregular. From $t = 50$~ms onward ($p_{41}$--$p_{53}$), {a local triangular-like
ordering develops in selected regions of the condensate, though both circulation
signs persist.} Hence, we do not refer to this as a perfect Abrikosov lattice (which
requires same-sign vortices). The dipolar interaction manifests as an elliptical
distortion and elongated core shapes. Vortex-antivortex annihilation is strongly
suppressed, allowing the mixed population to persist up to $100$~ms. 
Vortices are identified in the phase maps as points with $\pm 2\pi$ winding.

{In Case~1, the number of vortex pairs continues to grow up to $t \sim 40$~ms,
after which it remains roughly constant up to $t = 100$~ms, with small
fluctuations arising from the ongoing competition between vortex-pair creation
and annihilation under continuous forcing. This behavior is directly reflected in
the incompressible kinetic energy $E_K^{(i)}$ of Fig.~\ref{fig:ke_evolution}(a):
the curve grows sharply during the initial phase and begins to saturate around
$t \sim 40$~ms, remaining roughly saturated (though non-smooth) through the rest
of the simulation.} 

\subsection{Case 2: Obstacle removal at $t = 15$ ms}
\begin{figure}[htbp]
\centering
\includegraphics[width=\columnwidth]{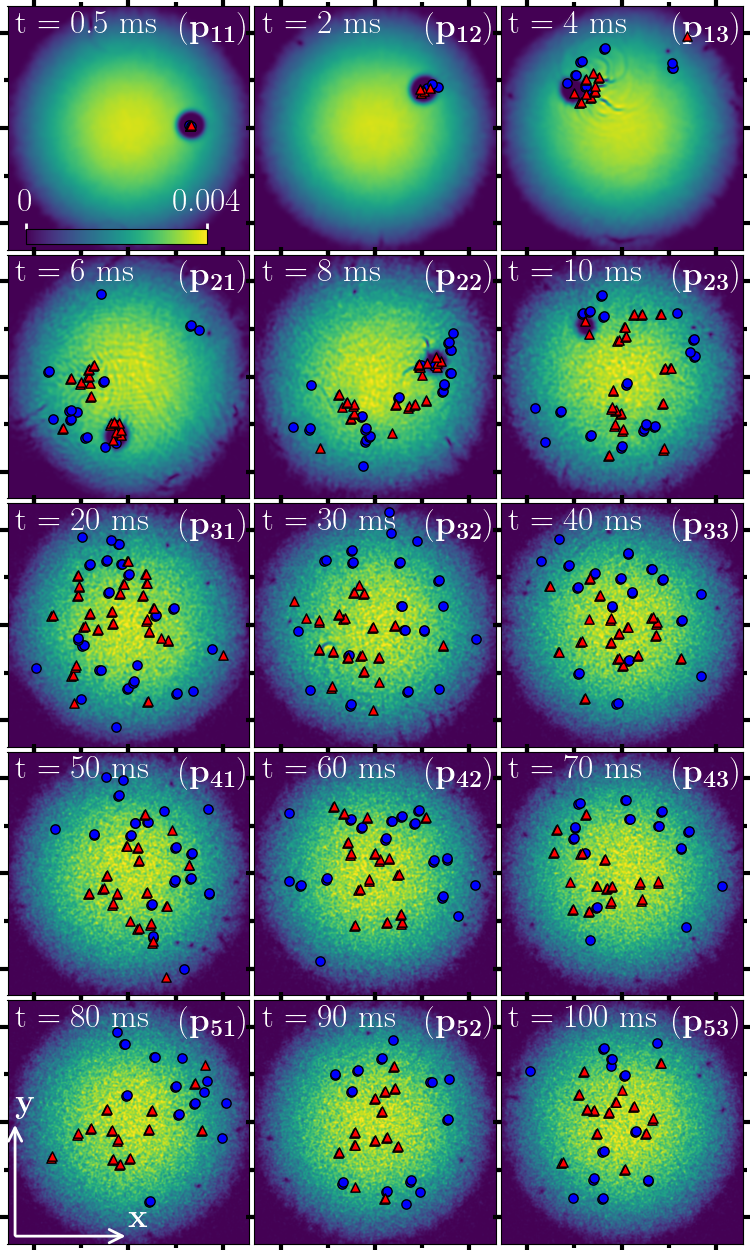}
\caption{
(Color online) Column-density evolution for a dipolar BEC in which the rotating
Gaussian obstacle is removed at $t = 15$~ms (Case 2). The amplitude ramps linearly,
$A(t)=A_0(1-t/15)$ for $t\le15$~ms, and is zero thereafter; all other parameters are
as in Fig.~\ref{fig:dens_case1}. {The first vortex pairs exit the obstacle at
$t \sim 3$~ms and proliferate during the stirring phase ($t\lesssim15$~ms).} After
obstacle removal, the vortex array evolves freely. {By $t\sim50$~ms, local
triangular-like ordering emerges in selected regions of the condensate, without
the formation of a globally ordered lattice.} No decay of the vortex population is
observed up to $100$~ms. {Insets at $t=40$~ms and $t=50$~ms show the vortex
positions with nearest-neighbor connectivity lines, highlighting the onset of
local sixfold-coordinated regions after obstacle removal.}}
\label{fig:dens_case2}
\end{figure}

\begin{figure}[htbp]
\centering
\includegraphics[width=\columnwidth]{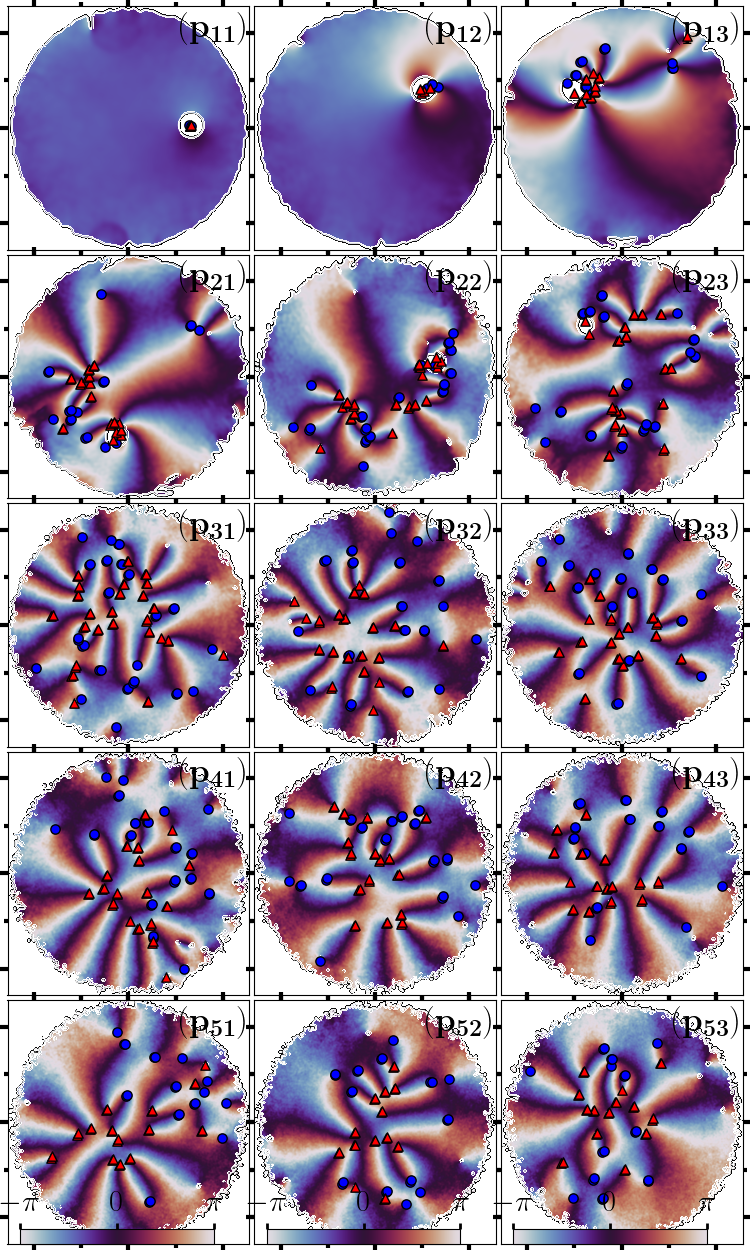}
\caption{
(Color online) Phase evolution corresponding to Fig.~\ref{fig:dens_case2} (Case 2). 
Phase singularities nucleate before $t=15$ ms and persist after obstacle removal. 
The free evolution allows the vortex ensemble to rearrange into an ordered lattice 
by $t\sim50$ ms, as confirmed by the regular arrangement of $2\pi$ phase dislocations.}
\label{fig:phase_case2}
\end{figure}

Figures~\ref{fig:dens_case2} and \ref{fig:phase_case2} show the evolution for Case 2.
The early dynamics ($t \le 10$~ms) resemble Case 1, but with slightly fewer vortices
due to the amplitude ramp-down. At $t = 4$~ms, the first vortex pairs exit the
obstacle {and become visible as dark density holes with $2\pi$ phase windings}.
At $t = 15$~ms, the amplitude reaches zero; no discontinuity appears in density or
phase, confirming adiabatic removal. {Between $t = 20$ and $40$~ms, additional
pairs separate and are resolved despite the absence of the obstacle, and after
$t \sim 50$~ms the vortex count stabilizes.} The final structure is a configuration
with {local triangular-like ordering in selected regions of the condensate},
cleaner than in Case 1 due to the absence of continuous forcing.

{The vortex population in Case~2 reaches its maximum around $t \sim 15$--$25$~ms
and then gradually decreases as the pairs relax in the absence of forcing. This
trend is mirrored by the incompressible kinetic energy $E_K^{(i)}$ in
Fig.~\ref{fig:ke_evolution}(b), which decreases monotonically after obstacle
removal, reflecting the relaxation and partial recombination of the vortex
ensemble.}


\subsection{Kinetic energies: Compressible-incompressible decomposition}
\label{sec:energy}

\begin{figure}[htbp]
\centering
\includegraphics[width=\columnwidth]{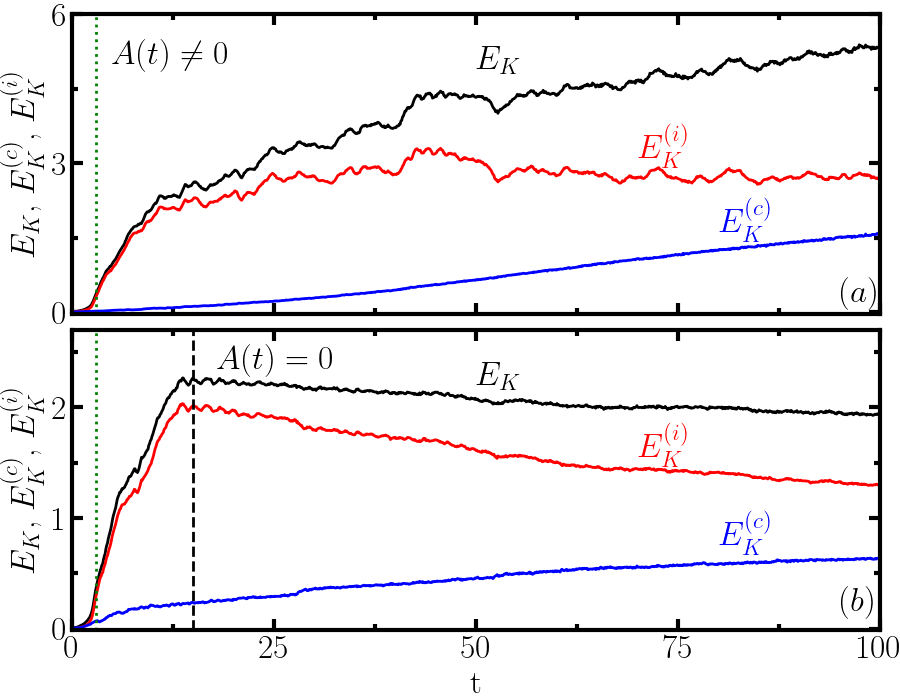}
\caption{
(Color online) Time evolution of the kinetic energy components for (a) Case 1 
(continuous stirring) and (b) Case 2 (obstacle removal at $t=15$ ms). 
Black solid curves: total kinetic energy ($E_K$). Blue dashed curves: 
compressible kinetic energy ($E_K^{(c)}$), associated with sound waves and phonons. 
Red dotted curves: incompressible kinetic energy ($E_K^{(i)}$), associated with 
vortex motion. The vertical dashed line at $t=15$ ms in panel (b) marks the 
obstacle removal time. The smooth angular velocity ramp ($t_{\text{acc}}=3$ ms) 
is active in both cases for $t<3$ ms.}
\label{fig:ke_evolution}
\end{figure}

From the scaled GP equation (Eq.~\ref{gpe_scaled}), which provides the time-dependent
density solution $n(\boldsymbol{\rho},t)$, the total energy of the system is given by
\begin{eqnarray}
E(t) &=& \int d^2\boldsymbol{\rho}\left[\frac{1}{2}|\nabla_{\rho}\psi|^2 + \frac{\rho^2}{2}+ 
V_G(\boldsymbol{\rho},t)n(\boldsymbol{\rho},t)\right] \nonumber\\
&+& \frac{g_{s}}{2} \int d^2\boldsymbol{\rho}\; n^2(\boldsymbol{\rho},t) + E_{dd},
\label{En}
\end{eqnarray}
where $E_{dd}$ accounts for the dipole-dipole interaction contribution. 
The total kinetic energy, defined as
\begin{equation}
E_K(t) \equiv \int d^2\boldsymbol{\rho}\left[ \frac{1}{2}|\nabla_{\rho}\psi|^2 \right],
\end{equation}
is the focus of this section. To analyze the nature of excitations, we introduce 
the current density $\mathbf{j}(\boldsymbol{\rho},t) = n(\boldsymbol{\rho},t) 
\mathbf{v}(\boldsymbol{\rho},t)$, which defines the velocity field

\begin{equation}
\mathbf{v}(\boldsymbol{\rho},t) = \frac{1}{2\mathrm{i}\, |\psi|^2} \left( \psi^\star \nabla_{\rho}\psi - \psi \nabla_{\rho} \psi^\star \right).
\label{current}
\end{equation}
The kinetic energy can then be expressed as
\begin{equation}
E_K(t) = \frac{1}{2} \int d^2\boldsymbol{\rho}\; n(\boldsymbol{\rho},t) \; |\mathbf{v}(\boldsymbol{\rho},t)|^2.
\label{Ken}
\end{equation}

To separate the energy contributions from compressible (sound wave) and incompressible
(vortex) motions, we define the density-weighted velocity field $\mathbf{u}
(\boldsymbol{\rho},t) \equiv \sqrt{n(\boldsymbol{\rho},t)} \, \mathbf{v}
(\boldsymbol{\rho},t)$. This field can be decomposed into an incompressible part 
$\mathbf{u}^{(i)}$ satisfying $\nabla_{\rho} \cdot \mathbf{u}^{(i)} = 0$, and 
a compressible part $\mathbf{u}^{(c)}$ satisfying $\nabla_{\rho} \times \mathbf{u}^{(c)} = 0$. 
With $\mathbf{u} = \mathbf{u}^{(i)} + \mathbf{u}^{(c)}$, the kinetic energy splits 
into two components:
\begin{equation}
E_K(t) = E_K^{(i)}(t) + E_K^{(c)}(t) \equiv \frac{1}{2} \int d^2\boldsymbol{\rho}\; \left( |\mathbf{u}^{(i)}|^2 + |\mathbf{u}^{(c)}|^2 \right).
\label{Ken-split}
\end{equation}

{Figure~\ref{fig:ke_evolution} presents the kinetic energy decomposition for both
cases. In both cases, all curves (total, compressible, and incompressible kinetic
energies) remain close to zero initially and begin to grow once the smooth angular
velocity ramp is ended ($t_{\text{acc}}=3$~ms).}
\begin{figure*}[htbp]
\centering
\includegraphics[width=\linewidth]{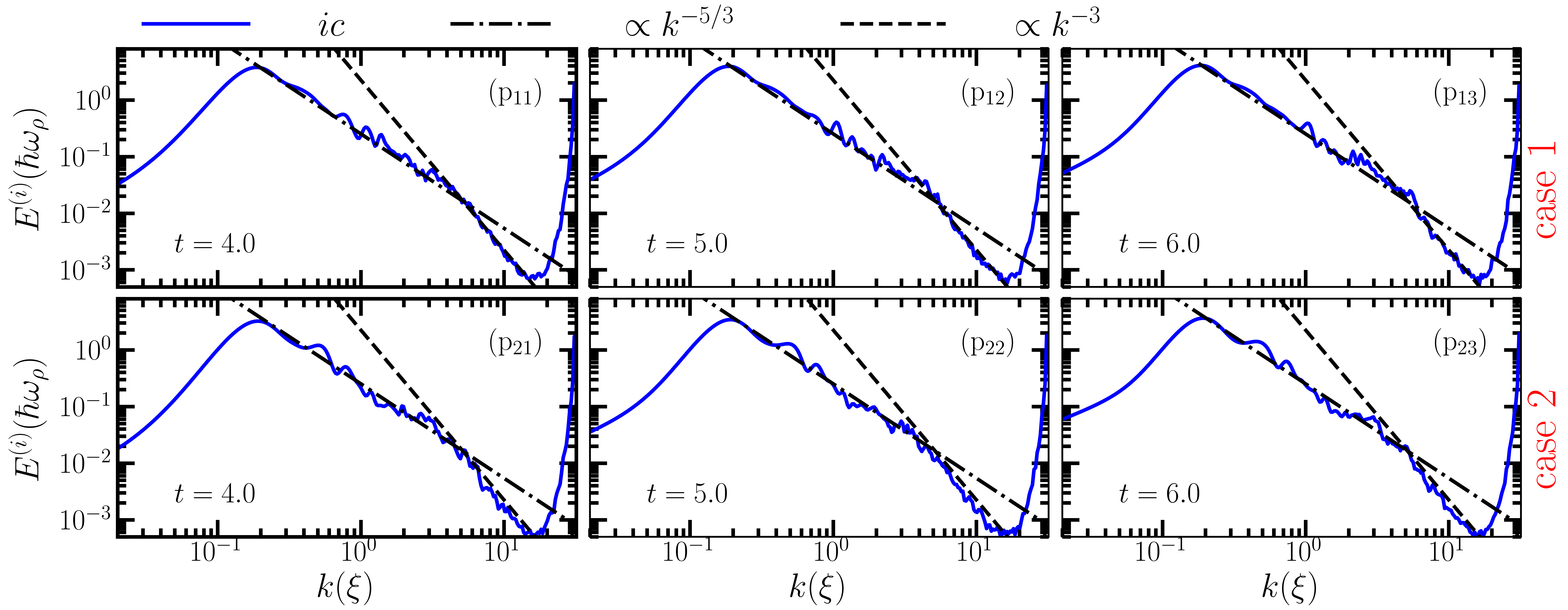}
\caption{
{(Color online)} 
These panels show the incompressible kinetic energy spectra, $E_i(k)$, at 
three different times, for 
Case 1 (upper row - continuous stirring) and 
Case 2 (lower row - obstacle removal). 
Left, middle, and right columns correspond to $t = 4$, $5$, and $6$ ms, respectively. 
As verified, the spectra show good agreement with the Kolmogorov 
scaling $k^{-5/3}$ (dot-dashed line) in the inertial range,
and $k^{-3}$ (dashed line) at the energy-containing range.
These results are indicating the emergence of a quasi-classical energy cascade 
even at these early times, before significant vortex proliferation. 
The agreement is slightly better for Case 2, where the absence of 
continuous forcing allows a cleaner cascade to develop.}
\label{fig:spec}
\end{figure*}
{For Case 1 (Fig.~\ref{fig:ke_evolution}a), there is a sudden increase in both the
total $E_K$ (black curve) and the incompressible component $E_K^{(i)}$ (red curve),
compared with the compressible component $E_K^{(c)}$ (blue curve). The blue curve is
smooth, whereas the black and red curves are non-smooth, exhibiting ups and downs
during their dynamical evolution. By $t=100$~ms, $E_K$, $E_K^{(i)}$, and $E_K^{(c)}$
increase from $0$ to $\approx 5.3$, $\approx 2.8$, and $\approx 1.5$ (in units of
$\hbar^2/(2m\ell_\rho^2)$), respectively. The black and blue curves continue to
increase, while the red curve shows signs of saturation and is itself non-smooth.}

{For Case 2 (Fig.~\ref{fig:ke_evolution}b), the energy evolution during stirring
($t < 15$~ms) is similar to Case 1 but with slightly lower values, owing to the
removal ramp. At $t = 15$~ms, no discontinuity appears in any component, confirming
adiabatic removal. Afterwards, $E_K$ and $E_K^{(i)}$ start to decrease, while
$E_K^{(c)}$ continues to increase smoothly. By $t=100$~ms, $E_K$, $E_K^{(i)}$, and
$E_K^{(c)}$ end up at $\approx 1.9$, $\approx 1.3$, and $\approx 0.6$ (in units of
$\hbar^2/(2m\ell_\rho^2)$), respectively. Comparing the two cases, the blue curve
$E_K^{(c)}$ is smooth in both and increases slowly up to the final time; the black
and red curves exhibit more pronounced ups and downs in panel (a) than in panel (b).}

{Two features stand out. First, the smooth monotonic growth of
$E_K^{(c)}$ in both panels confirms that the angular-velocity ramp
suppressed impulsive phonon generation; no acoustic burst accompanies
the onset of rotation. Second, the sharp jumps in $E_K^{(i)}$ for Case~1
track the successive nucleation and separation of vortex dipoles: the
curve begins to saturate around $t \sim 40$~ms and remains roughly
saturated through $t = 100$~ms, although it stays non-smooth throughout,
reflecting the ongoing rearrangement of the vortex ensemble under
continuous forcing. In Case~2, the monotonic decrease of $E_K^{(i)}$ after
obstacle removal reflects the relaxation of the vortex ensemble, while
the slow continued rise of $E_K^{(c)}$ indicates only weak residual sound
radiation.}

\subsection{Incompressible kinetic energy spectrum}

The incompressible kinetic energy spectrum $E_i(k)$ is computed from the Fourier
transform of the incompressible velocity field. It is related to the total 
incompressible kinetic energy by $E_K^{(i)} = \int E_i(k) \, dk$. Explicitly,
\begin{equation}
E_i(k,t) = \frac{1}{2} \int_0^{2\pi} |\tilde{\mathbf{v}}_i(\mathbf{k},t)|^2 \, k \, d\theta_k,
\end{equation}
where $\tilde{\mathbf{v}}_i(\mathbf{k},t)$ is the Fourier transform of $\mathbf{v}
_i(\boldsymbol{\rho},t)$. Figure~\ref{fig:spec} shows the incompressible 
kinetic energy spectrum $E_i(k)$ at early times ($t = 4$, $5$, and $6$ ms) 
for both cases. The spectrum is computed as $E_i(k) = \frac{1}{2} \int |\mathbf{v}
_i(\mathbf{k})|^2 \, k \, d\theta_k$, where $\mathbf{v}_i$ is the incompressible
component of the velocity field. For classical turbulence, the Kolmogorov spectrum 
$E(k) \propto k^{-5/3}$ characterizes the inertial range where energy cascades 
from large to small scales, while the $k^{-3}$ law is associated with the 
energy-containing range at larger scales.

At $t = 4$ ms (left column), vortices have just exited the obstacle. Both cases 
show a developing power-law behavior, though the range is limited due to the 
small number of vortices. By $t = 5$ ms (middle column), a clearer $k^{-5/3}$ 
range emerges, extending over approximately one decade in $k$. The $k^{-3}$ 
behavior at low $k$ is also visible. At $t = 6$ ms (right column), the 
spectra have further matured, with the $k^{-5/3}$ scaling becoming more pronounced. 
Case 2 (lower row) exhibits slightly cleaner power-law agreement than Case 1, 
likely because the amplitude ramp-down reduces perturbations that would 
otherwise distort the cascade.

The observation of these scaling laws at such early times ($t \sim 4$–$6$ ms) 
is noteworthy. In nondipolar condensates, the development of a $k^{-5/3}$ 
cascade typically requires longer evolution times and a larger vortex population. 
The rapid emergence of the cascade in the dipolar system may be attributed to 
the anisotropic long-range interactions, which facilitate energy transfer 
across scales. The persistence of the $k^{-5/3}$ scaling at later times (not shown)
correlates with the continued growth of the vortex population and the 
eventual formation of the partially ordered triangular array.

\section{Conclusion}
\label{sec4}
We have systematically studied vortex nucleation, proliferation, and self-organization
in a dipolar Bose-Einstein condensate stirred by a rotating Gaussian obstacle with 
a smooth angular velocity ramp ($t_{\text{acc}} = 3$~ms). Two stirring protocols 
were compared: continuous stirring with constant amplitude (Case 1), and obstacle 
removal at $t = 15$~ms with linear amplitude ramp-down (Case 2). In both cases, 
{the first vortex pairs exit the obstacle at $t \sim 3$~ms, becoming visible as
dark density holes with $2\pi$ phase windings in the panel at $t = 4$~ms, and vortex
dipoles subsequently separate into distinct phase singularities at $t \sim 8$--$10$~ms.}

Case 1 (continuous stirring) yields a growing vortex population that
{continues to increase up to $t \sim 40$~ms, after which it remains roughly
constant up to $t = 100$~ms, with small fluctuations arising from the ongoing
competition between vortex-pair creation and annihilation under continuous forcing.}
Case 2 (obstacle removal) yields {a vortex population that peaks around
$t \sim 15$--$25$~ms and then gradually decreases as the pairs relax in the absence
of forcing, stabilizing after $t = 50$~ms.} Remarkably, both protocols lead to
{local triangular-like ordering in selected regions of the condensate} by
$t \sim 50$~ms. We emphasize that this is not a perfect Abrikosov lattice, as both
vortices and antivortices coexist throughout the evolution. The dipolar interaction
manifests clearly in this structure through an elliptical distortion and elongated
vortex cores along the polarization axis.

A key finding is the strong suppression of vortex-antivortex annihilation in both
protocols. The mixed vortex-antivortex population remains stable up to $100$~ms 
with no observable decay, in stark contrast to nondipolar condensates where 
annihilation typically occurs within tens of milliseconds. This dipolar 
stabilization of vorticity is a central result of our study.

Kinetic energy decomposition confirms minimal phonon excitations, validating the
effectiveness of the smooth ramping protocol. {In Case~1, the compressible
component grows to $E_K^{(c)} \approx 1.5$ by $t = 100$~ms, while the incompressible
component, associated with vortex motion, rises to $E_K^{(i)} \approx 2.8$ and
begins to saturate around $t \sim 40$~ms, remaining roughly saturated (though
non-smooth) through the end of the simulation; this mirrors the saturation of the
vortex-pair count in the density panels. In Case~2, $E_K^{(c)} \approx 0.6$ and
$E_K^{(i)} \approx 1.3$ at $t = 100$~ms, with $E_K^{(i)}$ decreasing monotonically
after obstacle removal as the vortex ensemble relaxes, consistent with the
corresponding decrease in the vortex-pair count.} The incompressible kinetic energy
spectrum shows agreement with the Kolmogorov $k^{-5/3}$ scaling at early times
($t = 4$--$6$~ms), indicating the rapid emergence of a quasi-classical energy
cascade in the dipolar system.

These results establish the rotating Gaussian obstacle with smooth angular velocity
ramping as a controlled, minimally perturbative tool for vortex nucleation in dipolar
Bose-Einstein condensates. The observed partially ordered vortex arrays and the strong 
suppression of vortex annihilation open new avenues for investigating quantum 
turbulence in anisotropic superfluids. Future work should explore the role of the 
roton minimum in energy cascades, extend these studies to binary dipolar mixtures, 
and investigate the transition from partially ordered mixed-circulation arrays to 
same-sign Abrikosov lattices under stronger or longer continuous stirring.

\noindent {\bf Acknowledgements:}\\
S.S. thanks Funda\c c\~ao de Amparo \`a Pesquisa do Estado de S\~ao Paulo
(FAPESP) for the fellowship support, Proj.~2024/04174-8. 
L.T. also acknowledges partial support from FAPESP (Proj. 2024/01533-7) and
Conselho Nacional de Desenvolvimento Cient\'\i fico e Tecnol\'ogico
(Proc. 303263/2025-3). 

\bibliographystyle{amsplain}

\begin{thebibliography}{99}

\bibitem{Baranov2008} M.A. Baranov, {Theoretical progress 
in many-body physics with ultracold dipolar gases}, Phys. 
Rep. \textbf{464}, 71 (2008).

\bibitem{Lahaye2009} T. Lahaye, C. Menotti, L. Santos, M. Lewenstein, T. Pfau, 
{The physics of dipolar bosonic quantum gases}, Rep. Prog. Phys. \textbf{72}, 126401 (2009).

\bibitem{Griesmaier2006} A. Griesmaier, J. Stuhler, T. Koch, M. Fattori, T. Pfau, and 
S. Giovanazzi, {Comparing Contact and Dipolar Interactions in a Bose-Einstein Condensate}, 
Phys. Rev. Lett. \textbf{97}, 250402 (2006).

\bibitem{Lu2011} M. Lu, N.Q. Burdick, S.H. Youn, and B.L. Lev, {Strongly Dipolar 
Bose-Einstein Condensate of Dysprosium}, Phys. Rev. Lett. \textbf{107}, 190401 (2011).

\bibitem{Aikawa2012} K. Aikawa, A. Frisch, M. Mark, S. Baier, A. Rietzler, R. Grimm, 
and F. Ferlaino, {Bose-Einstein Condensation of Erbium}, Phys. Rev. Lett. \textbf{108}, 
210401 (2012).

\bibitem{2022Miyazawa}
Y. Miyazawa, R. Inoue, H. Matsui, G. Nomura, and M. Kozuma,
Bose-Einstein condensation of europium,
Phys. Rev. Lett. {\bf 129}, 223401 (2022).

\bibitem{1998Inouye} S. Inouye, M.R. Andrews, J. Stenger,  H.-J. Miesner, D.M. Stamper-Kurn, 
and W. Ketterle, {Observation of Feshbach resonances in a Bose-Einstein condensate}, 
Nature (London) {\bf 392}, 151 (1998).

\bibitem{Giovanazzi2002}S. Giovanazzi, A. G\"orlitz and T. Pfau, { Tuning the Dipolar 
Interaction in Quantum Gases}, Phys. Rev. Lett. \textbf{89}, 130401 (2002).

\bibitem{Sabari2015} 
S. Sabari, C.P. Jisha, K. Porsezian and V.A. Brazhnyi,
{Dynamical stability of dipolar Bose-Einstein condensates with temporal modulation of the s-wave scattering length}, 
Phys. Rev. E \textbf{92}, 032905 (2015).

\bibitem{Sabari2018a} 
S. Sabari and B. Dey,
{Stabilization of trapless dipolar Bose-Einstein condensates by temporal modulation of the contact interaction}, 
Phys. Rev. E \textbf{98}, 042203 (2018).

\bibitem{Sabari2022} S. Sabari, R.K. Kumar, R. Radha, and B. A. Malomed, 
Interplay between binary and three-Body interactions and enhancement of stability in trapless 
dipolar Bose–Einstein condensates. Appl. Sci. {\bf 12}, 1135 (2022).

\bibitem{2001Madison} 
K. W. Madison, F. Chevy, V. Bretin, and J. Dalibard,
Stationary States of a Rotating Bose-Einstein Condensate: Routes to Vortex Nucleation,
Phys. Rev. Lett. {\bf 86}, 4443

\bibitem{Tamil2019} R. Tamil Thiruvalluvar, S. Sabari, K. Porsezian, P. Muruganandam, 
{Vortex formation and vortex lattices in a Bose-Einstein condensate with Lee-Huang-Yang (LHY) 
correction}, Physica E {\bf 107}, 54 (2019).

\bibitem{Raman1999} C. Raman, M. K\"ohl, R. Onofrio, D. S. Durfee, C.E. Kuklewicz, 
Z. Hadzibabic, and W. Ketterle, { Evidence for a critical velocity in a Bose-Einstein 
condensed gas}, Phys. Rev. Lett. {\bf 83}, 2502 (1999). 

\bibitem{Raman2001} C. Raman, J.R. Abo-Shaeer, J.M. Vogels, K. Xu, and W. Ketterle, { 
Vortex Nucleation in a Stirred Bose-Einstein Condensate}, Phys. Rev. Lett. {\bf 87}, 210402 (2001). 

\bibitem{Henn2009} E.A.L. Henn, J.A. Seman, G. Roati, K.M.F. Magalhães, V.S. Bagnato, { Emergence of 
turbulence in an oscillating Bose-Einstein condensate}, Phys. Rev. Lett. {\bf 103}, 045301 (2009).

\bibitem{Lamporesi2013}G. Lamporesi, S. Donadello, S. Serafini, F. Dalfovo, and G. Ferrari, 
{ Spontaneous creation of Kibble–Zurek solitons in a Bose-Einstein condensate}, 
Nat. Phys. {\bf 9}, 656 (2013).

\bibitem{Kishor2025} R.K. Kumar, S. Sabari, A. Gammal, L. Tomio, { Rayleigh-Taylor, Kelvin-Helmholtz, and immiscible-to-miscible quenching instabilities in binary Bose-Einstein condensates}, Phys. Rev. A {\bf 112}, 033312 (2025).

\bibitem{Lauro2024} 
L. Tomio, A.N. da Silva, S. Sabari, and R.K. Kumar, {Dynamical Vortex Production and Quantum Turbulence in Perturbed Bose–Einstein Condensates}, Few-Body Sys. {\bf 65}, 13 (2024).

\bibitem{Lauro2026} 
L. Tomio, S. Sabari, A. Gammal, and R.K. Kumar, {Immiscible to Miscible Quenching Instabilities in Two-Dimensional Binary Bose-Einstein Condensates}, Few-Body Sys. {\bf 67}, 28 (2026).

\bibitem{Barenghi2001}C.F. Barenghi, R.J. Donnelly, W.F. Vinen, { Quantized Vortex Dynamics 
and Superfluid Turbulence}, eds. (Springer, New York), (2001).

\bibitem{Pethick2008}C. Pethick, H. Smith, {Bose-Einstein Condensation in Dilute Gases}, 
(Cambridge Univ Press, Cambridge, UK), 2nd Ed. (2008).

\bibitem{Sabari2017} S. Sabari, {Vortex formation and hidden vortices in dipolar Bose-Einstein 
condensates},  Phys. Lett. A \textbf{381}, 3062 (2017).

\bibitem{Sabari2026} 
S. Sabari, R. Radha, Lauro Tomio, B.A. Malomed,
{Stable quantum vortices in Lee-Huang-Yang dipolar superfluids}, Phys. Rev. A {\bf 113}, 023315 (2026).

\bibitem{Sabari2025} 
S. Sabari, R. Sasireka, R. Radha, A. Uthayakumar, and Lauro Tomio,
{Vortices in tunable dipolar Bose-Einstein condensates with attractive interactions}, Phys. Rev. A {\bf 111}, 053320 (2025).

\bibitem{Sabari2018} S. Sabari and R.K. Kumar, 
{Effect of an oscillating Gaussian obstacle in a dipolar Bose-Einstein condensate}, 
Eur. Phys. J. D \textbf{72}, 48 (2018).

\bibitem{Sabari2024cnf} 
S. Sabari, R.K. Kumar, and Lauro Tomio,
{Vortex-Antivortex Pair Production in Perturbed Dipolar Bose-Einstein Condensate}, Journal of Physics: Conference Series {\bf 2894}, 012015 (2024).

\bibitem{Sabari2024} 
S. Sabari, R.K. Kumar, and Lauro Tomio,
{Vortex dynamics and turbulence in dipolar Bose-Einstein condensates}, Phys. Rev. A {\bf 109}, 023313 (2024).

\bibitem{Sabari2018cnf} 
S. Sabari, {Vortices in a rotating dipolar Bose-Einstein condensate with two-and three-body interactions}, AIP Conference Proceedings {\bf 1942}, 030028 (2018).

\bibitem{Sabari2017cnf} 
S. Sabari and B. Dey, {Hydrodynamics of vortices and solitons in a dipolar Bose-Einstein condensate}, AIP Conference Proceedings {\bf 1832}, 030007 (2017).

\bibitem{Sabari2016cnf} 
S. Sabari, {Dynamics of vortices in a dipolar Bose-Einstein condensate with an oscillating potential},
Conference on Nonlinear Systems \& Dynamics, {\bf 16}, 18 (2016).

\bibitem{2022Klaus} L. Klaus, T. Bland, E. Poli, C. Politi, G. Lamporesi, E. Casotti, R. N. Bisset, 
M. J. Mark, and F. Ferlaino, { Observation of vortices and vortex stripes in a dipolar condensate}, 
Nat. Phys. {\bf 18}, 1453 (2022).

\bibitem{2026Zhu}  Y. Zhu, G. Krstulovic, and S. Nazarenko, {Strong and weak wave turbulence regimes in Bose–Einstein condensates}, Rep. Prog. Phys. {\bf 89}, 030501 (2026).

\bibitem{2026Mohapatra}
S. Mohapatra, A.J. Schaffer, P.G. Kevrekidis, R. Carretero-González, and B.P. Anderson, {Vortex comb: Eliminating vortices from Bose-Einstein condensates using optical lattices}, Phys. Rev. A {\bf 113}, 023305 (2026).

\bibitem{2025Jha} S.K. Jha, M.K. Verma, S.I. Mistakidis, and P.K. Mishra, {Energy spectra and fluxes of two-dimensional turbulent quantum droplets}, Phys. Rev. Fluids {\bf 10}, 064618 (2025).

\bibitem{2024Prasad} S.B. Prasad, N.G. Parker, and A.W. Baggaley, {Vortex-pair dynamics in three-dimensional homogeneous dipolar superfluids}, Phys. Rev. A {\bf 109}, 063323 (2024).
\end{thebibliography}

\end{document}